\documentclass[11pt, a4paper]{article}
\usepackage{fullpage}

\usepackage{belov_paper}

\title{Quantum Algorithm for Distribution-Free Junta Testing}
\author{
  Aleksandrs Belovs\thanks{Faculty of Computing, University of Latvia}
}

\usepackage{algorithm}

\begin{document}
\maketitle

\begin{abstract}
Inspired by a recent classical distribution-free junta tester by Chen, Liu, Serverdio, Sheng, and Xie (STOC'18), we construct a quantum tester for the same problem with complexity $O(k/\eps)$, which constitutes a quadratic improvement.

We also prove that there is no efficient quantum algorithm for this problem using quantum examples as opposed to quantum membership queries.

This result was obtained independently from the $\tO(k/\eps)$ algorithm for this problem by Bshouty.
\end{abstract}

\section{Introduction}
The steadily growing size of data calls for algorithms that work extremely fast: with linear or, preferably, sub-linear complexity.
To achieve this performance, certain assumptions must be made, usually in some sort of approximation guarantees.

Consider the problem of testing whether an object $f$ has some property $\cP$.
Typically, this task is hardest on instances $f$ that are close to the border: the ones that do not possess the property $\cP$ but are extremely close to doing so.
But often such meticulousness is not needed, as one can tolerate false positives that are close to true positives.  
What one usually wants is to filter out instances that are substantially far from having the property.
The framework of property testing\cite{rubinfeld:propertyTesting, goldreich:propertyTesting} does exactly this: distinguishes the objects having the property $\cP$ from the ones that are $\eps$-far from $\cP$.
This is an active area of research both for classical (randomised) and quantum testers.
See~\cite{montanaro:quantumProperyTest} for a survey on quantum property testing.

\paragraph{Junta Testing.}
Let us focus on the case of testing juntas, which is the topic of this paper.
A \emph{$k$-junta} is a Boolean function $h\colon\cube\to\bool$ that only depends on $k$ out of its $n$ input variables.  
In other words, $h$ is a $k$-junta iff there exists a subset $\{a_1,a_2,\dots,a_k\}\subseteq [n]$ such that $h(x_1,\dots,x_n) = \hat h(x_{a_1},x_{a_2},\dots,x_{a_k})$ for some function $\hat h\colon\bool^k\to\bool$.
A function $g\colon\cube\to\bool$ is \emph{$\eps$-far} from a $k$-junta iff $g$ differs from any $k$-junta in at least $\eps2^n$ points of the hypercube $\cube$.
A \emph{junta tester}, given access to a Boolean function $f$ and parameters $k$ and $\eps$, has to distinguish between the cases when $f$ is a $k$-junta and when $f$ is $\eps$-far from any $k$-junta.
It is usually assumed that $n$ is much larger than $k$, and the goal is to construct a tester whose complexity does not depend on $n$, and is as optimal in terms of $k$ and $\eps$ as possible.

\mytxtcommand{atici}{At{\i}c{\i}}

Junta testing has interesting history with developments in classical and quantum algorithms coming hand in hand. 
In 2002, the problem was considered by Fischer \etal~\cite{fischer:testingJuntas}, and an algorithm with query complexity $O\sA[(k\log k)^2/\eps]$ was constructed.
In 2007, a quantum algorithm was constructed by \atici and Servedio~\cite{atici:testingJuntas}, which uses $O(k/\eps)$ quantum examples.
But in 2009 classical algorithms caught up with an $O(k/\eps + k \log k)$-query algorithm by Blais~\cite{blais:testingJuntas}.
This is optimal due the lower bound by Sa\u glam~\cite{saglam:heat} (see also~\cite{chockler:testingJuntasLower}).
In 2015, however, quantum complexity of the problem was improved to $O(\sqrt{k/\eps}\log k)$ queries by Ambainis \etal~\cite{belovs:gappedGroupTesting}.
This was shown to be almost optimal by Bun \etal~\cite{bun:polynomialStrikesBack}.
Thus, query complexity of junta testing is well-understood both classically and quantumly.

\paragraph{Distribution-free testing.}
The distribution-free property testing model was introduced by Goldreich \etal in~\cite{goldreich:propertyTesting}.
It is similar to the usual model of property testing, but the distance to the property $\cP$ is measured with respect to some unknown distribution $\cD$ over the hypercube $\cube$.
That is, $g$ is $\eps$-far from a $k$-junta with respect to $\cD$, iff $\Pr_{x\sim\cD} [g(x)\ne h(x)]\ge\eps$ for any $k$-junta $h$.
The tester, in addition to oracle queries to $f$, can sample from $\cD$.
The tester should work for any distribution $\cD$, and the complexity measure is the worst-case sum of the number of queries to $f$ and samples from $\cD$.
Thus, distribution-free property testing is at least as hard as usual property testing.

The motivation behind distribution-free property testing model is similar to that of the PAC learning model~\cite{valiant:PAC}.
The uniform distribution might not be the right one to measure the distance, and it might be hard to get to know what the relevant probability distribution is.
Thus, we would like to make as few assumptions on the distribution as possible.  
But the algorithm must have some access to the distribution in order to solve the problem, and sampling is one of the weakest modes of access.

\paragraph{Results.}
Recently, Liu \etal~\cite{chen:distributionFreeJuntas} constructed a distribution-free randomised junta tester with complexity $\tO(k^2)/\eps$.
Our main result is a distribution-free \emph{quantum} junta tester with a (slightly better than) quadratic improvement in $k$.

\begin{thm}
\label{thm:main}
There exists an algorithm that, given quantum membership oracle access to a Boolean function $f\colon\cube\to\bool$, classical sample access to a probability distribution $\cD$ on $\cube$, and parameters $k$ and $\eps$, performs the following task.
If $f$ is a $k$-junta, it accepts with probability 1.
If $f$ is $\eps$-far from any $k$-junta with respect to $\cD$, it rejects with probability at least $1/2$.
The algorithm uses $O(k/\eps)$ queries to $f$ and samples from $\cD$.
\end{thm}

Up to our knowledge, this is the first distribution-free quantum property tester.
We allow quantum membership queries to $f$.
Additionally, we allow classical (not quantum!) sampling from $\cD$.
If one allows quantum example oracle access to $\cD$, it is possible to get quadratic improvement in terms of $\eps$ as well, see \rf{sec:discussion}.
We give additional justification to our model in \rf{sec:uselessness}, where we show that quantum example oracle is of little use to solve this problem.

Our algorithm is inspired by the algorithm by Liu \etal, and it not only has smaller complexity, but is also conceptually simpler that the classical algorithm.
Similarly to the tester by \atici and Servedio, the improvement stems from the fact that quantum algorithms can efficiently Fourier sample.
Actually, the only quantum subroutine we use is the ability to Fourier sample the input function $f$ restricted to arbitrary hypercube of $\cube$.

\paragraph{Related Results.}
During the review phase of this paper, classical complexity of the problem was improved to $\tO(k/\eps)$ queries by Bshouty~\cite{bshouty:distributionFreeJunta}, which is optimal up to logarithmic factors.

\section{Preliminaries}
\label{sec:prelim}
We use notation $[n]=\{1,2,\dots,n\}$.  For $x\in\cube$ and $T\subseteq [n]$, $x^T$ stands for the string $x$ with the bits in $T$ flipped.  We write $x^i$ instead of $x^{\{i\}}$ for $i\in[n]$.

The following ways of accessing the input function $f\colon\cube\to\bool$ are used in the paper.
A \emph{classical example} is a pair $(x,f(x))$, where $x$ is drawn from some probability distribution (or uniformly if no distribution is specified).  
A \emph{classical membership oracle} is a black-box that on a query $x$ returns the value of $f(x)$.
A \emph{quantum example} is a quantum state of the form $\sum_{x\in\cube} \sqrt{\cD_x} \ket|x>\ket|f(x)>$ for some probability distribution $\cD$.
Again, if no distribution is specified, we assume the uniform one.
A \emph{quantum example oracle} is a quantum subroutine that performs the transformation $\ket|0>\mapsto\sum_{x\in\cube} \sqrt{\cD_x} \ket|x>\ket|f(x)>$.
Quantum example oracles are more powerful than quantum examples: for instance, one can use quantum amplitude amplification with them.
Finally, a \emph{quantum membership oracle} is a quantum subroutine performing the transformation $\ket|x> \mapsto (-1)^{f(x)}\ket|x>$ for all $x\in\cube$.
It is the same as the usual quantum input oracle.

In order to distinguish quantum example and membership oracles, we use terms quantum \emph{sample} and \emph{query}, respectively, for their execution.
Classically, we use terms ``sample'' and ``example'' as synonymous.
For each quantum oracle, we also assume access to its inverse.

The main technical tool we use is Fourier sampling.
The \emph{Fourier transform} (also known as the Walsh-Hadamard transform) of a Boolean function $f$ is the linear mapping
\begin{equation}
\label{eqn:Fourier}
\frac{1}{\sqrt{2^n}} \sum_{x\in\cube} (-1)^{f(x)} \ket|x> \stackrel{H^{\otimes n}}{\longmapsto} 
\sum_{S\subseteq [n]} \hat f(S) \ket |S>,
\end{equation}
where $\hat f(z)$ are the \emph{Fourier coefficients} of $f$.  Here the subset $S$ on the right-hand side can be identified with its characteristic string.

The Fourier transformation can be efficiently implemented on a quantum computer: it only requires $n$ Hadamard gates $H = \frac{1}{\sqrt 2}\begin{pmatrix}1&1\\1&-1\end{pmatrix}$.
If the state on the right-hand side of~\rf{eqn:Fourier} is measured, the outcome $S$ is obtained with probability $\hat f(S)^2$.  This is known as \emph{Fourier sampling}.  The sum of $\hat f(S)^2$ equals 1 since $H^{\otimes n}$ is a unitary transformation.

We only use the following two properties of the Fourier transform.  
First, if $\hat f(S)\ne 0$, then $f$ depends on all variables in $S$.
Second, 
\begin{equation}
\label{eqn:FourierEmpty}
\hat f(\emptyset) = \frac 1{2^n} \sum_{x\in\cube} (-1)^{f(x)}.
\end{equation}

\section{The Algorithm}
\label{sec:main}
In this section we describe our algorithm and prove \rf{thm:main}.
Before we proceed with this task, let us introduce some notation.
Everywhere in this section we assume that an input function $f\colon\cube\to\bool$ is fixed.

A variable $i\in [n]$ is called \emph{relevant} if there exists an input $x\in\cube$ such that $f(x) \ne f(x^i)$.  
Note that a function $f$ is a $k$-junta if and only if it has at most $k$ relevant variables.

A \emph{cube} $B=(x,y)$ is a subcube of the hypercube $\cube$ and is specified by its two opposite vertices $x$ and $y$.
That is, $B$ is the set of all bit-strings in $\cube$ that agree to either $x$ or $y$ in each position.
Let $I(B)$ be the set of variables where $x$ and $y$ disagree.
A cube $B=(x,y)$ is called \emph{relevant} if $f(x)\ne f(y)$.
Note that if a cube $B$ is relevant, then at least one variable in $I(B)$ is relevant.
\medskip

\begin{algorithm}[p]
\caption{Quantum Distribution-Free Junta Testing}
\label{alg:main}
\enumstart
\item Let $S\gets\emptyset$, $\cB\gets\emptyset$
\item Repeat while $|S|+|\cB|\le k$, but no more than $18k$ times:
\enumstart
\negmedskip
\item If $\cB$ is empty, $B\gets \mbox{GenerateCube($S$)}$.  If $B$ is not fail, let $\cB\gets\{B\}$.  Continue with the next iteration of the loop.
\item Otherwise, let $B=(x,y)$ be any cube in $\cB$.
\item Let $T\gets \mbox{FourierSample($B$)}$.  If $T\ne\emptyset$, remove $B$ from $\cB$, let $S\gets S\cup T$, and continue with the next iteration of the loop.
\item Otherwise, generate a uniformly random subset $T\subseteq I(B)$.  Let $z = x^T$ and $t = y^T$.
\item If $f(z) = f(t) = f(y)$, then remove $B$ from $\cB$, add the cubes $(x,z)$ and $(x,t)$ to $\cB$, and continue with the next iteration of the loop.
\item If $f(z) = f(t) = f(x)$, then remove $B$ from $\cB$, add the cubes $(z,y)$ and $(t,y)$ to $\cB$, and continue with the next iteration of the loop.
\enumend
\item If $|S|+|\cB|> k$, reject.  Otherwise, accept.
\enumend
\end{algorithm}

\begin{algorithm}[p]
\caption{GenerateCube($S$) subroutine, classical version}
\label{alg:genCube}
\begin{enumerate}
\item Repeat $2/\eps$ times:
\begin{enumerate}
\negmedskip
\item Sample $x$ from $\cD$.  Let $T$ be a uniformly random subset of $[n]\setminus S$.
\item If $f(x) \ne f(x^T)$ return the cube $(x,x^T)$.
\end{enumerate}
\item Return `fail'.
\end{enumerate}
\end{algorithm}

\begin{algorithm}[p]
\caption{FourierSample($B=(x,y)$)}
\label{alg:Fourier}
\begin{enumerate}
\item Prepare the state $\ket|x>=\ket|x_1>\ket|x_2>\cdots\ket|x_n>$ on $n$ qubits.
\item Apply the Hadamard operator $H$ to the qubits in $I(B)$, and get the state $\frac1{\sqrt{|B|}}\sum_{z\in B} \ket|z>$.
\item Apply the quantum membership oracle and obtain the state $\frac1{\sqrt{|B|}}\sum_{z\in B} (-1)^{f(z)}\ket|z>$.
\item Apply the Hadamard operator $H$ to the qubits in $I(B)$ and measure them.  Return the set $T$ formed by the bits where the measurement outcome is 1.
\end{enumerate}
\end{algorithm}

A formal description of the tester is given in \rf{alg:main}, and it depends on two subroutines: GenerateCube and FourierSample given in Algorithms~\ref{alg:genCube} and~\ref{alg:Fourier}, respectively.
The subroutine GenerateCube($S$) finds a cube $B$ such that $I(B)$ does not intersect $S$.
The subroutine FourierSample($B$) performs Fourier sampling from the function $f$ restricted to the subcube $B$.

The algorithm maintains a subset of variables $S\subseteq [n]$ and a collection of cubes $\cB$.
The following invariants are maintained throughout the algorithm:
\begin{equation}
\label{eqn:inv}
\parbox{.9\textwidth}
{
\itemstart
\item All variables in $S$ and all cubes in $\cB$ are relevant.
\item The set $S$ and all $I(B)$, as $B$ ranges over $\cB$, are pairwise disjoint.
\itemend
}
\end{equation}

\begin{clm}
The invariants in~\rf{eqn:inv} are maintained throughout the algorithm.
\end{clm}

\pfstart
Clearly, the invariants are satisfied at the beginning of the algorithm when both $S$ and $\cB$ are empty.
Consider all the steps of \rf{alg:main} where $S$ or $\cB$ change.
In 2(a), $B$ is a relevant cube and $I(B)$ does not intersect $S$ by the requirement on the subroutine GenerateCube.
In 2(c), the set $T$ returned by FourierSample has non-zero Fourier coefficient in $f$ restricted to $B$.  This means that all the variables in $T$ are relevant.  Also $T\subseteq I(B)$, hence, the new $S$ does not intersect any of the remaining cubes in $\cB$.
In 2(e), we have $f(z)\ne f(x)$ and $f(t)\ne f(x)$, hence the two new cubes are relevant.
Also note that $t=x^{I(B)\setminus T}$.  Hence, $I(x,z)$ and $I(x,t)$ are disjoint and do not intersect $S$ nor the remaining cubes in $\cB$.
The case 2(f) is similar.
\pfend

Concerning the terminating condition in Steps 2 and 3 of \rf{alg:main}, we have the following result.

\begin{prp}
\label{prp:ending}
If $S$ and $\cB$ satisfy the conditions in~\rf{eqn:inv} and $|S| + |\cB| > k$, then $f$ is not a $k$-junta.
\end{prp}

\pfstart
Each cube $B\in\cB$ contains a relevant variable.
By the second condition in~\rf{eqn:inv}, all these variables are distinct and different from the variables in $S$.
Thus, together with the variables in $S$, $f$ has more than $k$ relevant variables.
This means that $f$ is not a $k$-junta.
\pfend

Concerning the GenerateCube subroutine, we have the following lemma from~\cite{chen:distributionFreeJuntas}.
\begin{lem}[\cite{chen:distributionFreeJuntas}, Lemma 3.2. arXiv version]
\label{lem:main}
Assume $f\colon\cube\to\bool$ is $\eps$-far from any $k$-junta with respect to $\cD$, and $S$ is a subset of $[n]$ of size at most $k$.
Then,
\[
\Pr_{x\sim\cD,\; T\subseteq [n]\setminus S} \skA[f(x)\ne f(x^T)] \ge \eps/2,
\]
where $x$ is sampled from $\cD$, and $T$ is a uniformly random subset of $[n]\setminus S$.
\end{lem}

\begin{cor}
\label{cor:genCube}
In the assumptions of \rf{lem:main}, the probability that the GenerateCube($S$) subroutine returns `fail' is at most $1/2$.
\end{cor}

\pfstart
By \rf{lem:main}, the probability that the subroutine returns `fail' is at most $(1-\eps/2)^{2/\eps} < 1/\ee < 1/2$.
\pfend

The number of iterations of the loop in Step 2 of \rf{alg:main} is based on the following lemma.

\begin{lem}
\label{lem:growth}
Assume $f$ is $\eps$-far from a $k$-junta relative to $\cD$, and consider one iteration of the loop in Step 2 of \rf{alg:main}.
Let $S$ and $\cB$ be the values of these variables before the iteration, and $S'$ and $\cB'$ after the iteration.
If $|S|\le k$, then, with probability at least 1/3, we have $2|S'|+|\cB'| \ge 2|S|+|\cB|+1$.
\end{lem}

\pfstart
Assume first $\cB=\emptyset$.  Then by \rf{cor:genCube}, with probability at least $1/2$, the GenerateCube subroutine does not fail and the size of $\cB$ becomes 1, which gives $2|S'|+|\cB'| = 2|S|+|\cB|+1$.

Now consider the case when $\cB$ is not empty, and let $B$ be the cube selected on step 2(b).
Denote by $f|_B$ the function $f$ restricted to the inputs in the subcube $B$.
There are two cases: $f|_B$ is $1/3$-far from a constant function, or $f|_B$ is $1/3$-close to a constant function relative to the uniform distribution on $B$.

In the first case, by~\rf{eqn:FourierEmpty}, the absolute value of the Fourier coefficient of $\emptyset$ is at most $2/3-1/3=1/3$.  Hence, with probability $8/9$, the set $T$ in 2(c) is non-empty.  
In this case, the size of $\cB$ is reduced by 1, but the size of $S$ grows by at least 1, which gives $2|S'|+|\cB'| \ge 2|S|+|\cB|+1$.

In the second case, when $f$ is $1/3$-close to a constant function, the probability that $f(z)=f(t)$ is at least $1/3$.  In this case, the size of $S$ does not change, but the size of $\cB$ grows by 1, which gives $2|S'|+|\cB'| = 2|S|+|\cB|+1$.
\pfend

\pfstart[Proof of \rf{thm:main}]
Now we can prove the theorem.
If $f$ is a $k$-junta, \rf{alg:main} always accepts due to \rf{prp:ending}.
Let us prove that if $f$ is $\eps$-far from a $k$-junta relative to $\cD$, then the algorithm rejects with probability at least $1/2$.

Assume for a moment there is no upper bound of $18k$ on the number of iterations of the loop in Step 2 of the algorithm.
Let $P(i)$ denote the number of iterations of the loop after which it holds that $2|S|+|\cB|\ge i$ or $|S|+|\cB|>k$.
By \rf{lem:growth}, we have that $\bE[P(i+1)-P(i)] \le 3$.
As $P(0)=0$, by linearity of expectation, we have that $\bE[P(3k)] \le 9k$.
By Markov's inequality, 
\begin{equation}
\label{eqn:markov}
\Pr[P(3k)\ge 18k]\le 1/2.
\end{equation}
Note that $2|S|+|\cB|\ge 3k$ implies $|S|+|\cB|>k$.
Thus, \rf{eqn:markov} means that \rf{alg:main} rejects with probability at least $1/2$.

The complexity of the algorithm is $O(k/\eps)$, as it performs $O(k)$ iterations of the loop, and each iteration costs at most $O(1/\eps)$. 
\pfend

\section{Uselessness of Quantum Examples}
\label{sec:uselessness}
One interesting feature of the quantum junta tester by \atici and Servedio~\cite{atici:testingJuntas} is that it only uses quantum examples and not quantum or classical membership queries.
This constitutes an exponential improvement since exponentially many classical examples are required to solve the problem (see \rf{lem:classicalExamples} below).
Also, this is still the best known quantum algorithm that only uses quantum examples, as the algorithm by Ambainis \etal~\cite{belovs:gappedGroupTesting} uses quantum membership queries.

Our algorithm in~\rf{sec:main} uses quantum membership queries, and a natural question arises whether it is possible to attain similar complexity using only quantum examples.
We show that this is impossible:
contrary to the uniform case, in the distribution-free case, exponentially many quantum examples are required.
As it is unclear whether quantum examples should come from the uniform distribution or from $\cD$, we show that neither of them works.

\begin{thm}
\label{thm:samples}
Assume a quantum algorithm has quantum example oracle access to a Boolean function $f$ with respect to both uniform probability distribution and $\cD$.
Then, $2^{\Omega(k)}$ executions of these oracles are required to distinguish whether $f$ is a $k$-junta or $\Omega(1)$-far away from any $k$-junta with respect to $\cD$.
\end{thm}

Recall that quantum example oracles with respect to the uniform distribution and $\cD$ are quantum subroutines that perform the following transformations
\[
\ket|0>\mapsto \frac1{\sqrt{2^n}} \sum_{x\in\cube} \ket|x>\ket|f(x)>,
\qqand
\ket|0>\mapsto \sum_{x\in\cube} \sqrt{\cD_x} \ket|x>\ket|f(x)>,
\]
respectively.
In the remaining part of this section, we sketch the proof of \rf{thm:samples}.

We start with proving that classical examples do not help to solve the problem even in the uniform case.
Interestingly, the following result does not appear explicitly in prior publications, however, it is tacitly assumed in a number of papers.
We add a simple proof of this lemma for completeness.

\begin{lem}
\label{lem:classicalExamples}
In the uniform model, $\Omega(2^{k/2})$ classical examples are required to distinguish whether a given Boolean function $f\colon\cube\to\bool$ is a $k$-junta or is $\Omega(1)$-far from any $(n-1)$-junta.
\end{lem}

\pfstart
Consider the following two probability distributions on Boolean functions $f\colon\cube\to\bool$.
$\cH$ is the uniform probability distribution on the functions $h(x_1,\dots,x_n)$ that only depend on the variables $x_1,\dots,x_k$.
$\cG$ is the uniform probability distribution on all Boolean functions $f$.
Each function in the support of $\cH$ is a $k$-junta, while a function $g\sim\cG$ is $\Omega(1)$-far from any $(n-1)$-junta with high probability~\cite[Lemma 3.1]{chockler:testingJuntasLower}.
Hence, the tester should be able to distinguish these two probability distribution with bounded error.

Let $x^{(1)},\dots,x^{(t)}$ be the examples obtained by the tester.
If $t = o(2^{k/2})$, then, with high probability, every two inputs in this sequence disagree on their first $k$ bits.
But if this is the case, then the values $f(x_1),\dots,f(x_t)$ form a uniformly random string in $\bool^t$ regardless whether $f$ is sampled from $\cH$ or from $\cG$.
Hence, the tester cannot distinguish these two probability distributions with bounded error.
\pfend

Let $\cH$ denote the set of functions that only depend on $x_1,\dots,x_k$ and $\cG$ denote the set of functions that are $\Omega(1)$-far from any $(n-1)$-junta.
\rf{lem:classicalExamples} shows that exponentially many classical examples are needed to distinguish these two classes.
However, due to the algorithm by \atici and Servedio, it is possible to efficiently distinguish them using quantum examples.
We proceed by modifying the classes $\cH$ and $\cG$ twice.
First, we make the uniform quantum example oracle useless, and then the quantum example oracle with respect to $\cD$.

We start with uniform quantum examples.
The idea is to restrict the probability distribution $\cD$ to a small subcube so that uniform examples do not give much information on the inputs in $\cD$.
Let $x_{[k]}$ denote the substring $(x_1,\dots,x_k)\in\bool^k$ of $x$.
Define $\cH'$ and $\cG'$ as classes of Boolean functions $f'\colon\bool^{n+k}\to\bool$ on the variables $(x,y) = (x_1,\dots,x_n,y_1,\dots,y_k)$ defined in the following way.
If $f$ comes from $\cH$ (respectively, $\cG$), then the corresponding function $f'$ in $\cH'$ (respectively, $\cG'$) is defined by
\[
f'(x_1,\dots,x_n,y_{1},\dots,y_{k}) =
\begin{cases}
f(x_1,\dots,x_n),&\text{if $y = x_{[k]}$;}\\
0,&\text{otherwise}.
\end{cases}
\]
Let $\cD'$ be the uniform probability distribution on the strings $(x,y)\in\bool^{n+k}$ satisfying $y = x_{[k]}$.
Any function in $\cH'$ is a $2k$-junta, and any function from $\cG'$ is $\Omega(1)$-far from any $(n-1)$-junta with respect to $\cD'$, and, hence, $\Omega(1)$-far from any $2k$-junta if $n>2k$.

Uniform quantum examples are useless here.
Indeed, the uniform quantum example is $2^{-\Omega(k)}$-close to the quantum example corresponding to the all-0 function:
\begin{equation}
\label{eqn:zero}
\ket|0>\mapsto \frac1{\sqrt{2^{n+k}}} \sum_{x\in\cube}\sum_{y\in\bool^k} \ket|x>\ket|y>\ket|0>.
\end{equation}
Hence, we can replace the uniform quantum example oracle corresponding to a function $f'$ in $\cH'\cup\cG'$ with~\rf{eqn:zero} and it will not significantly affect the output of the algorithm unless it makes $2^{\Omega(k)}$ samples.
\medskip

Our next goal is to get rid of the quantum example oracle relative to $\cD$.
The idea is to scatter the distribution $\cD$ so that it is impossible to make use of interference, and the quantum example oracle becomes essentially equivalent to the classical example oracle.
Let $\cH''$ and $\cG''$ be classes of Boolean functions $f''\colon\bool^{2n+k}\to\bool$ obtained from $\cH'$ and $\cG'$, respectively, by extending them with $n$ irrelevant variables $z_1,\dots,z_{n}$.
For $\pi\colon\cube\to\cube$ a permutation,
we define a distribution $\cD''_\pi$ as follows.
The distribution $\cD''_\pi$ is over strings $(x,y,z)$, where $x$ is a uniformly random string from $\cube$, $y=x_{[k]}$ and $z=\pi(x)$.
Clearly, all the functions in $\cH''$ are $(2k)$-juntas and the functions in $\cH''$ are $\Omega(1)$-far from any $(n-1)$-junta with respect to $\cD''_\pi$.
The uniform quantum samples are still not useful by the same argument as above.

Let us now prove that quantum samples from $\cD''_\pi$ are also useless in this situation.
Consider the quantum sampler from $\cD''_\pi$.  It acts as follows
\begin{equation}
\label{eqn:xOracle}
\ket|0>\mapsto \frac1{\sqrt{2^{n}}} \sum_{x\in\cube} \ketA|x>\ketA|x_{[k]}>\ketA|\pi(x)>\ketA|f(x)>,
\end{equation}
where $f$ is from $\cH$ or $\cG$.
But note that this state can be obtained in one query to the quantum oracle
\begin{equation}
\label{eqn:yOracle}
\ket|z>\ket|0>\mapsto \ket|z>\ketA|\pi^{-1}(z)>\ketA|f\sA[\pi^{-1}(z)]>,
\end{equation}
where we denoted $z = \pi(x)$ and permuted the registers.
Hence, the complexity of the tester using the oracle in~\rf{eqn:xOracle} is at least the complexity of the tester using the oracle in~\rf{eqn:yOracle},
which is the standard quantum oracle corresponding to a function
\[
z \mapsto \sB[{\pi^{-1}(z), f\sA[\pi^{-1}(z)] }].
\]

Note that the problem of testing juntas using the oracle from~\rf{eqn:yOracle} is symmetric with respect to the permutations of $z$ because the permutation $\pi$ can be arbitrary.
By \cite{chailloux:symmetricFunctions}, a quantum algorithm for a symmetric function can obtain at most a cubic improvement compared to a randomised algorithm with access to the oracle.
However, for a random $\pi$, access to this oracle is equivalent to uniform classical examples of the function $f$.
Hence, the algorithm requires $2^{\Omega(k)}$ samples by~\rf{lem:classicalExamples}.

\section{Discussion}
\label{sec:discussion}
We have constructed a quantum algorithm with complexity $O(k/\eps)$, which gives a quadratic improvement in terms of $k$ when compared to~\cite{chen:distributionFreeJuntas}.
In the algorithm, we assume we only have classical access to $\cD$.
It is also possible to have a quantum sampler from $\cD$, that is, an oracle of the form
\[
\ket|0> \mapsto \sum_{x\in\cube} \sqrt{\cD_x} \ket|x>\ket|\psi_x>,
\]
where $\cD_x$ is the probability of $x$ in $\cD$, and $\psi_x$ are some arbitrary unknown normalized quantum states.
(Note that because of the unknown $\psi_x$, this is a weaker model of access than the quantum samplers we defined in \rf{sec:prelim}.)
In this case, it is possible to apply quantum amplitude amplification to the GenerateCube subroutine.
The complexity of the subroutine becomes $O(1/\sqrt{\eps})$, and the complexity of the whole algorithm becomes $O(k/\sqrt{\eps})$.

So far, it is unclear whether the complexity of our quantum algorithm.
In principle, it is not excluded that there exists a quantum algorithm with complexity $\tO(\sqrt{k/\eps})$.

\section*{Acknowledgements}
I am thankful to Srinivasan Arunachalam and Ronald de Wolf for helpful discussions about this problem.

This research is partly supported by the ERDF grant number 1.1.1.2/VIAA/1/16/113.

\bibliographystyle{habbrvM}
\bibliography{belov}

\begin{thebibliography}{10}

\bibitem{belovs:gappedGroupTesting}
A.~Ambainis, A.~Belovs, O.~Regev, and R.~de~Wolf.
\newblock {Efficient Quantum Algorithms for (Gapped) Group Testing and Junta
  Testing}.
\newblock In {\em Proc.\ of 27th ACM-SIAM SODA}, pages
  \myhref{http://dx.doi.org/10.1137/1.9781611974331.ch65}{903--922}, 2016.
\newblock  \myhref{http://arxiv.org/abs/1507.03126}{{\ttfamily
  arXiv:1507.03126}}.

\bibitem{atici:testingJuntas}
A.~At{\i}c{\i} and R.~A. Servedio.
\newblock Quantum algorithms for learning and testing juntas.
\newblock {\em Quantum Information Processing},
  6(5):\myhref{http://dx.doi.org/10.1007/s11128-007-0061-6}{323--348}, 2007.
\newblock  \myhref{http://arxiv.org/abs/0707.3479}{{\ttfamily
  arXiv:0707.3479}}.

\bibitem{blais:testingJuntas}
E.~Blais.
\newblock Testing juntas nearly optimally.
\newblock In {\em Proc.\ of 41st ACM STOC}, pages
  \myhref{http://dx.doi.org/10.1145/1536414.1536437}{151--158}, 2009.

\bibitem{bshouty:distributionFreeJunta}
N.~H. Bshouty.
\newblock Almost optimal distribution-free junta testing.
\newblock  \myhref{http://arxiv.org/abs/1901.00717}{{\ttfamily
  arXiv:1901.00717}}, 2019.

\bibitem{bun:polynomialStrikesBack}
M.~Bun, R.~Kothari, and J.~Thaler.
\newblock {The polynomial method strikes back: Tight quantum query bounds via
  dual polynomials}.
\newblock In {\em Proc.\ of 50th ACM STOC}, pages
  \myhref{http://dx.doi.org/10.1145/3188745.3188784}{297--310}, 2018.
\newblock  \myhref{http://arxiv.org/abs/1710.09079}{{\ttfamily
  arXiv:1710.09079}}.

\bibitem{chailloux:symmetricFunctions}
A.~Chailloux.
\newblock A note on the quantum query complexity of permutation symmetric
  functions.
\newblock In {\em Proc.\ of 10th ACM ITCS}, volume 124 of {\em LIPIcs}, pages
  \myhref{http://dx.doi.org/10.4230/LIPIcs.ITCS.2019.19}{19:1--19:7}. Dagstuhl,
  2019.
\newblock  \myhref{http://arxiv.org/abs/1810.01790}{{\ttfamily
  arXiv:1810.01790}}.

\bibitem{chockler:testingJuntasLower}
H.~Chockler and D.~Gutfreund.
\newblock A lower bound for testing juntas.
\newblock {\em Information Processing Letters},
  90(6):\myhref{http://dx.doi.org/10.1016/j.ipl.2004.01.023}{301--305}, 2004.

\bibitem{fischer:testingJuntas}
E.~Fischer, G.~Kindler, D.~Ron, S.~Safra, and A.~Samorodnitsky.
\newblock Testing juntas.
\newblock {\em Journal of Computer and System Sciences},
  68(4):\myhref{http://dx.doi.org/10.1016/j.jcss.2003.11.004}{753--787}, 2004.
\newblock Earlier: \myhref{http://dx.doi.org/10.1109/SFCS.2002.1181887}{{\em
  FOCS'02}}.

\bibitem{goldreich:propertyTesting}
O.~Goldreich, S.~Goldwasser, and D.~Ron.
\newblock Property testing and its connection to learning and approximation.
\newblock {\em Journal of the ACM},
  45(4):\myhref{http://dx.doi.org/10.1145/285055.285060}{653--750}, 1998.
\newblock Earlier: \myhref{http://dx.doi.org/10.1109/SFCS.1996.548493}{{\em
  FOCS'96}}.

\bibitem{chen:distributionFreeJuntas}
Z.~Liu, X.~Chen, R.~A. Servedio, Y.~Sheng, and J.~Xie.
\newblock Distribution-free junta testing.
\newblock In {\em Proc.\ of 50th ACM STOC}, pages
  \myhref{http://dx.doi.org/10.1145/3188745.3188842}{749--759}, 2018.
\newblock  \myhref{http://arxiv.org/abs/1802.04859}{{\ttfamily
  arXiv:1802.04859}}.

\bibitem{montanaro:quantumProperyTest}
A.~Montanaro and R.~de~Wolf.
\newblock A survey of quantum property testing.
\newblock {\em Theory of Computing Graduate Surveys},
  7:\myhref{http://dx.doi.org/10.4086/toc.gs.2016.007}{1--81}, 2016.
\newblock  \myhref{http://arxiv.org/abs/1310.2035}{{\ttfamily
  arXiv:1310.2035}}.

\bibitem{rubinfeld:propertyTesting}
R.~Rubinfeld and M.~Sudan.
\newblock Robust characterizations of polynomials with applications to program
  testing.
\newblock {\em SIAM Journal on Computing},
  25(2):\myhref{http://dx.doi.org/10.1137/S0097539793255151}{252--271}, 1996.

\bibitem{saglam:heat}
M.~Sa{\u{g}}lam.
\newblock Near log-convexity of measured heat in (discrete) time and
  consequences.
\newblock In {\em Proc.\ of 59 th IEEE FOCS}, pages
  \myhref{http://dx.doi.org/10.1109/FOCS.2018.00095}{967--978}, 2018.
\newblock  \myhref{http://arxiv.org/abs/1808.06717}{{\ttfamily
  arXiv:1808.06717}}.

\bibitem{valiant:PAC}
L.~G. Valiant.
\newblock A theory of the learnable.
\newblock {\em Communications of the ACM},
  27(11):\myhref{http://dx.doi.org/10.1145/1968.1972}{1134--1142}, 1984.

\end{thebibliography}

\end{document}